\documentclass[usenatbib,letters]{mnras}
\usepackage[dvips]{graphicx}
\usepackage[english]{babel}
\usepackage{amsmath}
\usepackage{amssymb}
\usepackage{newtxtext,newtxmath}
\usepackage{float}

\newcommand{\EQ}{\begin{equation}}
\newcommand{\EN}{\end{equation}}
\newcommand{\EQA}{\begin{eqnarray}}
\newcommand{\ENA}{\end{eqnarray}}

\newcommand{\bfl}{\mbox{\boldmath $l$} {}}
\newcommand{\kk}{\mbox{\boldmath $k$} {}}
\newcommand{\BB}{\mbox{\boldmath $B$} {}}
\newcommand{\FF}{\mbox{\boldmath $F$} {}}
\newcommand{\mkG}{\,{\rm \mu G}}

\newcommand{\meanrho}{\overline{\rho}}

\newcommand{\sigmarm}{\bar{\sigma}_{\rm RM}}

\def\Rm{\rm Rm}

\def\Rey{{\rm Re}}
\def\Pm{\rm Pm}

\title[Rotation measure in young galaxies]{Faraday rotation signatures of fluctuation dynamos in
young galaxies}
\author[Sur, Bhat \& Subramanian]{Sharanya Sur$^{1}$\thanks{E-mail: sharanya.sur@iiap.res.in (SS); 
pbhat@mit.edu (PB); kandu@iucaa.in (KS)} 
Pallavi Bhat$^{2}$, \& Kandaswamy Subramanian$^{3}$\\
$^1$Indian Institute of Astrophysics, 2nd Block, Koramangala, Bangalore 560034, India\\
$^2$Plasma Science and Fusion Center, MIT, Cambridge, MA 02139, USA\\
$^3$IUCAA, Post Bag 4, Ganeshkhind, Pune 411007, India}

\begin{document}

\pagerange{\pageref{firstpage}--\pageref{lastpage}} \pubyear{2017}

\maketitle

\label{firstpage}

\begin{abstract} 
Observations of Faraday rotation through high-redshift galaxies have revealed that they host
coherent magnetic fields that are of comparable strengths to those observed in nearby 
galaxies. These fields could be generated by fluctuation dynamos. We use idealized numerical 
simulations of such dynamos in forced compressible turbulence up to rms Mach number of $2.4$ 
to probe the resulting rotation measure (RM) and the degree of coherence of the magnetic field. 
We obtain rms values of RM at dynamo saturation of the order of 45 - 55 per cent of the value 
expected in a model where fields are assumed to be coherent on the forcing scale of turbulence.
We show that the dominant contribution to the RM in subsonic and transonic cases comes from 
the general sea of volume filling fields, rather than from the rarer structures. However, in the 
supersonic case, strong field regions as well as moderately overdense regions contribute 
significantly. Our results can account for the observed RMs in young galaxies.
\end{abstract}

\begin{keywords}
dynamo -- magnetic fields -- MHD -- turbulence -- galaxies : high-redshift
\end{keywords}

\section{Introduction}

Mg II absorption systems probed by \citet{Bernet+08}, \citet{Bernet+10}, \citet{Farnes+14} and 
\citet{MCS17} reveal the existence of ordered $\mkG$ strength magnetic fields in young galaxies at $z\sim 1$ 
when the Universe was $\sim 6\,{\rm Gyr}$ old. The question of how such ordered fields of strengths comparable 
to those found in present-day galaxies are generated at early epoch remains an open question. It is well 
known that weak initial seed magnetic fields embedded in a conducting fluid can be amplified by a 
three dimensional random or turbulent flow, provided the magnetic Reynolds number ($\Rm$) is above 
a critical instability threshold. This process referred to as {\it Fluctuation} or {\it Small-scale} dynamo, amplifies 
seed magnetic fields exponentially fast (on eddy-turnover time-scales) by random stretching of field lines 
by the turbulent eddies, until some saturation process sets in \citep{K68,ZRS90,S99,HBD04,Scheko+04,
BS05,Cho+09,Fed+11,BSS12,BS13,Fed16}. Apart from generating and maintaining fields in galaxy 
clusters, fluctuation dynamo can also be a suitable candidate for generating magnetic fields in young 
galaxies at high redshifts \citep{SSK13, Roy+16, RT17}. This is mainly due to the following reasons. 
First, the excitation of the dynamo only requires $\Rm$ to be larger than 
a modest critical value $\sim 100$, in comparison to more special conditions (like the presence of 
differential rotation and helicity of the flow) necessary to turn on the large-scale dynamo. 
Moreover, fluctuation dynamos amplify fields on time-scales much shorter than large-scale 
dynamos where fields can only grow and order themselves on time scales of a few times $\sim 10^{8}\,{\rm yr}$. 
They can also potentially lead to sufficiently coherent fields to explain the observations \citep[][hereafter BS13]{BS13}. 
Lastly, recent high resolution simulations of halo collapse shows them to be capable of driving rapid field 
amplification even before disk formation (at $z \sim 25-30$) by tapping into the energy of turbulent 
motions in the halo gas \citep{Sur+10, Latif+13}. Taken together, these benefits potentially assist the 
fluctuation dynamo to generate the first fields in galaxies. It is then natural to ask, if the saturated fields 
generated by fluctuation dynamos in young galaxies are coherent enough and the extent to which the 
Faraday rotation measure (RM) obtained from such fields compares with the observational estimates 
from Mg II absorption systems. Moreover, how sensitive is the RM to regions of different fields strengths 
and densities? Addressing these key issues forms the subject of this Letter. 

Given the complex nature of the problem, we adopt a numerical approach as our main tool. To this effect, 
we focus on simulating fluctuation dynamos in periodic domains with artificial turbulent driving. 
In subsonic flows, previous works encompassing a wide range of viscosity and magnetic resistivity values 
have shown that the magnetic correlation length is much larger than the resistive scale \citep{HBD03,HBD04} 
leading to a significant contribution to the RM \citep[][BS13]{SSH06, CR09}. However, turbulence in 
high-redshift galaxies is driven by a combination of star formation feedback and gravitational instabilities 
from cold gas accretion flows \citep{BD03, Imm+04, CDB10, GDC12, Agertz+15}. Coupled with 
efficient cooling this results in supersonic turbulence \citep{Greif+08}. Thus, ascertaining the degree of 
coherence of the field and the RM calls for extending the simulations to the compressible regime involving 
transonic and supersonic flows. Such efforts have only recently become possible 
\citep[e.g.,][]{HBM04,Fed+11, Gent+13, SPS14a, SPS14b, Fed16, YCK16}. Motivated by the success 
of these efforts, we focus on fluctuation dynamo action in transonic and supersonic turbulent flows. 

The Letter is organized as follows. Section~\ref{sims} outlines our numerical setup and initial conditions. 
In Section~\ref{timevol}, we first describe the time evolution of the magnetic energy and the power spectra, 
followed by a visual impression of the nonlinear saturated state of the dynamo. Analysis of the temporal 
evolution of the RM, its dependence on regions with different field strengths and densities as well as the 
evolution of the velocity and magnetic integral scales are presented in Section~\ref{rmeasure}. The last 
section presents a summary of our results and conclusions. 

\section{Numerical setup and initial conditions} \label{sims}

To address our objectives, we analyse the data from simulations of randomly forced, non-helical, statistically 
homogenous and isotropic small-scale dynamo action. These simulations were performed with the FLASH 
code \citep{Fryxell+00} (version 4.2). The initial setup of our simulations is similar to the ones described in 
detail in an earlier paper by \citet{SPS14a, SPS14b}. In the following, we only highlight the essential 
features. 

We adopt an isothermal equation of state with initial density and sound speed set to unity in a box of unit length 
having $512^{3}$ grid points with periodic boundary conditions. 
Turbulence is driven artificially as a stochastic Ornstein - Uhlenbeck process \citep{EP88, Benzi+08}, with 
a finite time correlation. Note that both ISM and intracluster medium (ICM) turbulence are expected to be a mix 
of solenoidal (sheared) and compressive modes \citep{ES04, FKS08, PJR15}. However, in order to maximize 
the efficiency of the small-scale dynamo, our artificial forcing comprises only of solenoidal modes 
(i.e., $\nabla\cdot\FF =0$). Except run A where turbulence is forced in the wavenumber range 
$1\leq |\kk|L/2\pi \leq 2$ (to compare with a similar run from BS13), all other runs were forced in the 
wavenumber range  $1\leq |\kk|L/2\pi \leq 3$, such that the average forcing wavenumber was 
$k_{\rm f}\,L/2\pi \sim 2$, where $L$ is the length of the box. The amplitude of the turbulent forcing was varied 
to obtain runs with rms Mach numbers ($\mathcal{M}$) ranging from subsonic to supersonic. We further 
initialized our setup with two different choices of the initial magnetic field, consisting of either - 
${\BB} = B_{0}[0,0,1]$ or ${\BB} = B_{0}[0,0,\sin(2\pi x)]$. 
The strength of the initial field is varied by varying the plasma beta parameter 
$\beta = p_{\rm th}/p_{\rm m}$ in the range $10^{7} - 10^{9}$. Here $p_{\rm th}$ and $p_{\rm m}$ are the 
thermal and the magnetic pressures, respectively. Moreover, we used the un-split staggered-mesh algorithm 
in FLASH with a constrained transport to maintain $\nabla\cdot\BB$ to machine precision \citep{LD09, Lee13} 
and the HLLD Riemann solver \citep{MK05}, instead of artificial viscosity to capture shocks. 
%%%%%%%%%%%%%%%%%%%%%%%%%%%%%%%%%%%%%%%%%%%%%%%%%%%%
\begin{table}
\centering
\caption{Summary of the simulation parameters. From the left to right : Run name, initial $\beta$, 
rms Mach number, magnetic Prandtl number ($\Pm = \Rm/\Rey$) and the fluid Reynolds number.
Runs A and D are initialized with $B_{z} = B_{0}$, while runs B and C start with $B_{z} = B_{0}\sin(2\pi x)$.
All runs have resolutions of $512^{3}$.
}
\vspace{-0.8em}
\resizebox{0.45\textwidth}{!}{%
\begin{tabular}{|c|c|c|c|c|} \hline 
Run &  $\beta_{\rm in}$ & $\mathcal{M}_{\rm rms}$ & $\Pm$ & ${\rm Re} = u\;l_{\rm f}/\nu$ \\ \hline 
A & $10^{7}$ & $\approx 0.1$ & 1 & $\approx 800$  \\ 
B & $2.5\times10^{9}$ & $\approx 0.3$ & $1$ & $\approx 1250$ \\ 
C & $2.5\times10^{7}$ & $\approx 1.1$ & 1 & $\approx 1250$ \\ 
D & $10^{7}$ & $\approx 2.4$ & ----- & ------  \\ \hline 
\end{tabular}
}
\label{sumsim}
\vspace{-1em}
\end{table}
%%%%%%%%%%%%%%%%%%%%%%%%%%%%%%%%%%%%%%%%%%%%%%%%%%%%%%%%

Table~\ref{sumsim} provides a summary of the parameters of all the runs considered in this study. These 
consist of three non-ideal runs at $\Pm = \Rm/\Rey = \nu/\eta = 1$, with $\mathcal{M} = 0.1, 0.3,$ 
and $1.1$ and one ideal run at $\mathcal{M} = 2.4$. Since there is no explicit viscosity and 
magnetic resistivity in ideal MHD, kinetic and magnetic energies in run D are dissipated by 
numerical diffusion, which is expected to occur at the grid resolution scale. 

\section{Time evolution and field structure}
\label{timevol}

\begin{figure}
\centering
\includegraphics[width=0.98\columnwidth]{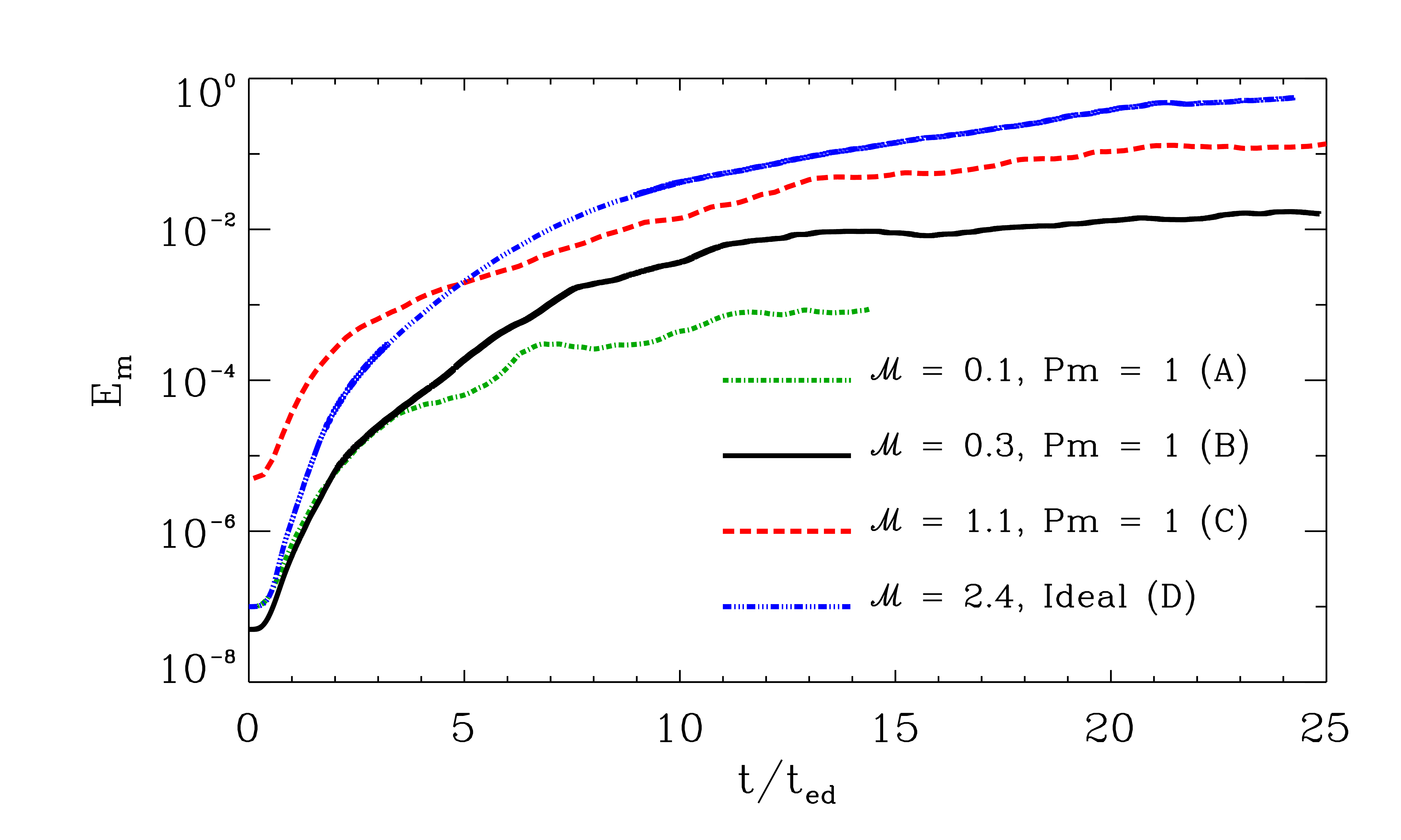}
\vspace{-1em}
\caption{Evolution of magnetic energy as a function of $t/t_{\rm ed}$.  
}
\label{evolmag}
\vspace{-1.5em}
\end{figure}

Turbulent motions randomly stretch and fold the weak initial magnetic field in all the different runs, 
leading to the emergence of growing random fields [see for e.g. fig.~2 of \citet{SPS14b}]. 
These random fields grow rapidly and ultimately saturate due to the influence of the Lorentz force.
Fig.~\ref{evolmag} shows the evolution of the magnetic energy ($E_{\rm m}$) as a function of time 
(expressed in units of the eddy-turnover time $t_{\rm ed}$). Similar to earlier studies on small-scale 
dynamo growth \citep[e.g.][]{Cho+09, PJR15}, the evolution of $E_{\rm m}$ consists of an initial 
exponential growth phase lasting up to $\sim(2-4\,t_{\rm ed})$, followed by an intermediate phase, 
eventually saturating within $10 - 30$ percent of the equipartition value. 
%%%%%%%%%%%%%%%%%%%%%%%%%%%%%%%%%%%%%%%%%%%%%%%%%%%%%%%%
%%%%%%%%%%%%%%%%%%%%%%%%%%%%%%%%%%%%%%%%%%%%%%%%%%%%%%
%%%%%%%%%%%%%%%%%%%% Power spectra evolution %%%%%%%%%%%%%%%%%%%%%%%
\begin{figure}
\begin{center}
\includegraphics[width=1.0\columnwidth,height=0.15\textheight]{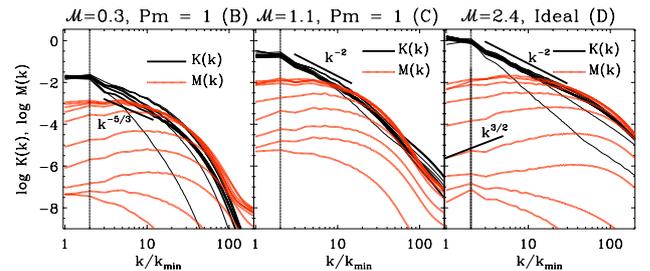}
\vspace{-1em}
\caption{Time evolution of the kinetic and magnetic energy spectra for runs : B - first column, C - second 
column, and D - third column. The dotted vertical line in each panel shows the turbulent forcing wavenumber. 
The wavenumber is normalized in units of $k_{\rm min} = 2\pi$.
\label{spectra1}}
\end{center}
\vspace{-1.5em}
\end{figure}
%%%%%%%%%%%%%%%%%%%%%%%%%%%%%%%%%%%%%%%%%%%%%%%%%%%%%%%%

In Fig.~\ref{spectra1}, we compare the evolution of the kinetic and magnetic energy spectra, $K(k,t)$ and 
$M(k,t)$, as a function of $k/k_{\rm min}$ for three runs : B (subsonic), C (transonic), and D (supersonic), 
respectively. 
$K(k,t)$ is depicted by thin (growth phase) and thick (saturated) solid lines, while $M(k,t)$ is shown as red 
dashed lines.
At late times $K(k,t)$ exhibits the expected Kolmogorov $k^{-5/3}$ scaling for run B, while for runs C and D, 
the slope is $\sim k^{-2}$, typical for supersonic turbulence. The comparison of $M(k,t)$ reveals a few interesting 
features. The magnetic spectra $M(k)$ is initially peaked at $k \sim 2$ in all three runs as the large scale field is 
tangled and twisted on the scale of the flow.
Because of the lack of explicit viscosity and resistivity, run D can accommodate a larger range of wave numbers 
for the inertial range of turbulence and for the growth of the small-scale fields. Thus, the well known self-similar 
evolution of $M(k)$, with a peak at large $k$, in the growth phase is seen more pronouncedly in run D compared 
to runs B and C. Specifically, 
the figure shows that the peak of $M(k)$ shifts from $k \sim 1 $ to $15$ in run B and to $k\sim 10$ in run C. In 
comparison, $M(k)$ in run D peaks at extremely tiny scales $k \sim 50-60$ by $t = 6\,t_{\rm ed}$. Once 
saturation is achieved at larger $k$, all three runs show that the peak of $M(k)$ shifts towards smaller wave 
numbers $k \sim 4-6$, by the time the dynamo saturates. 

\begin{figure}
\begin{center}
\includegraphics[width=1.0\columnwidth]{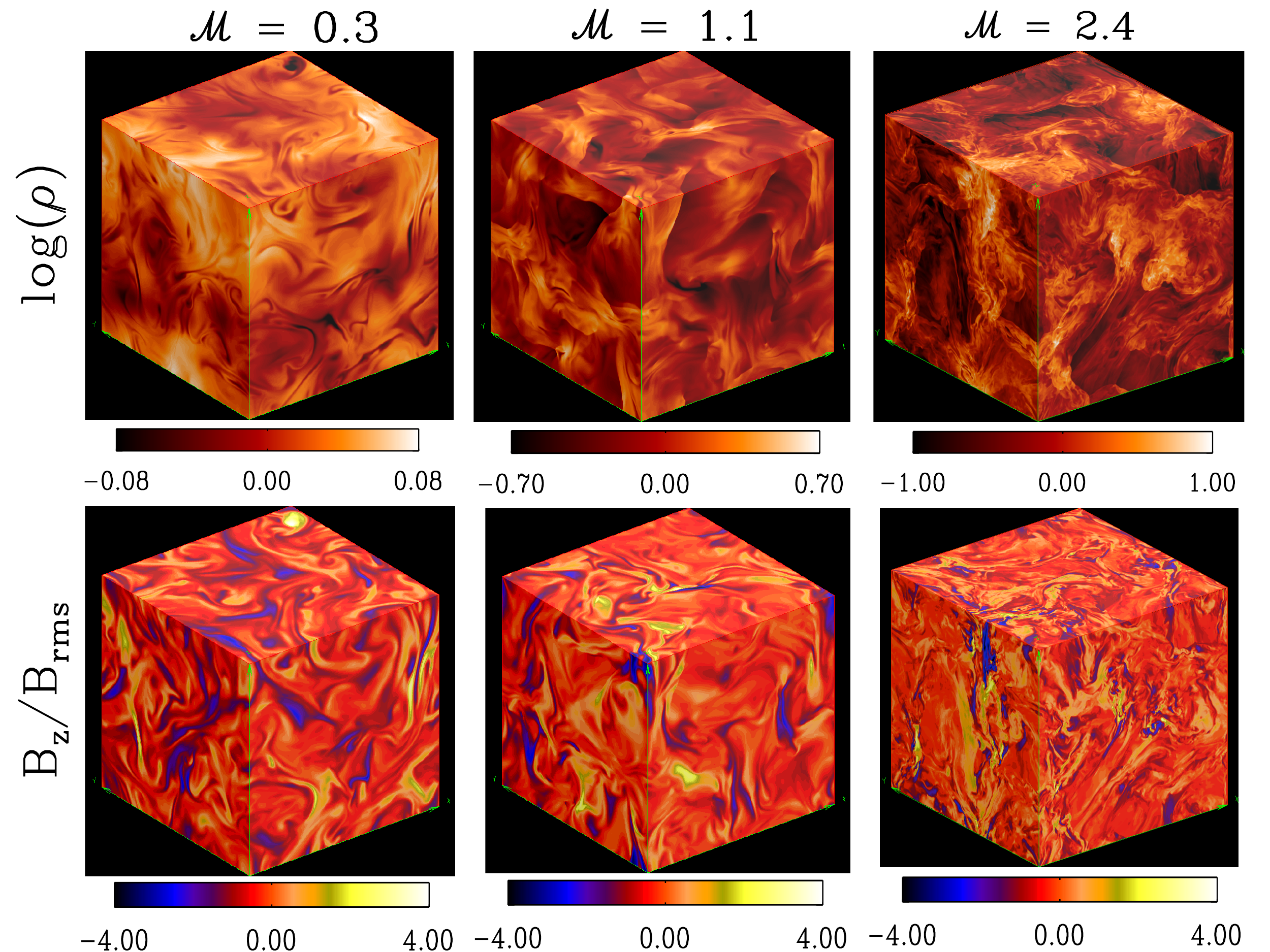}
\vspace{-1em} 
\caption{Three-dimensional volume renderings of the logarithm of the density (upper row) and the 
$z$-component of the magnetic field normalized by the rms value (lower row) in the saturated phase for 
runs : B (left), C (middle), and D (right) corresponding $\mathcal{M} = 0.3, 1.1$ and $2.4$, respectively. 
From left to right, the snapshots are shown at the following times : $t/t_{\rm ed} = 24.8, 23$ and $24$, 
respectively.
\label{volrend_sat}}
\end{center}
\vspace{-1.5em} 
\end{figure}

In Fig.~\ref{volrend_sat}, we show the three-dimensional volume rendering of the logarithm of 
the density (top row) and the $z$ component of the magnetic field normalized to the rms value (bottom row)
on the periphery of the box for the above mentioned runs, in the saturated phase. Apart from revealing evidence 
of turbulent stretching, the density structures are also sensitive to the compressibility of the flow. 
The magnetic field structures appear to be coherent on much larger scales, compared to the kinematic phase. 
We also find the field to be more intermittent and less space filling in the ideal run in comparison to the non-ideal 
runs. This enhanced intermittency is reminiscent of high $\Pm$ runs \citep[see fig. 5.7 of][]{BS05}, 
but could also arise due to the compression of magnetic fields in to high density regions. 
%%%%%%%%%%%%%%%%%%%% Saturated - Vol rendering %%%%%%%%%%%%%%%%%%%%%%%
% Use * for making the figure in single column
%%%%%%%%%%%%%%%%%%%%%%%%%%%%%%%%%%%%%%%%%%%%%%%%%%%%%%%%

\section{Faraday Rotation from $3N^{2}$ lines of sight}
\label{rmeasure}

Faraday rotation is a powerful tool to obtain information about the line-of-sight (LOS) magnetic field. 
The magnetized plasma in young galaxies together with charged particles causes a rotation 
of the polarization angle of linearly polarized emission from background radio sources. The observed 
polarization angle is altered by an amount proportional to the square of the observing wavelength. 
The constant of proportionality, the rotation measure  ${\rm RM} = K\int_{L}\,n_{\rm e}\BB\cdot d{\bfl}$. 
Here, $n_{\rm e}$ is the thermal electron density, $\BB$ is the magnetic field, 
$K = 0.81\,{\rm rad\,m^{-2}\,cm^{3}\,\mkG^{-1}\,pc^{-1}}$ is a constant and the integration is along 
the LOS '$L$' from the source to the observer. As our simulations include transonic and supersonic flows 
that results in significant density fluctuations along the LOS, we retain $n_{\rm e}$ inside the integral for 
all the runs. Following the methodology outlined in \citet{SSH06} and BS13, we directly compute 
$\int{\rho\BB\cdot d\bfl}$ for each of the different 
runs listed in Table~\ref{sumsim} and hence the RM over $3N^{2}$ LOS, along each of 
the $x, y$, and $z$ directions. For example, if the LOS integration is along $z$, the RM at a given 
location $(x_{i}, y_{i})$ is expressed as a discrete sum of $B_{z}$, 
\EQ
{\rm RM} (x_{i}, y_{i}, t) = \frac{K}{\mu m_{p}} \sum_{j=0}^{N-1} 
\left(\frac{L}{N}\right)\,\rho B_{z}\left(x_{i}, y_{i}, \frac{L}{N}j, t\right). 
\label{rmeq}
\EN
Here, $n_{\rm e} = \rho/\mu m_{p}$ is expressed in terms of the density, $L$ is the length of the box,  
and $N$ is the number of grid points. Magnetic fields generated by the fluctuation dynamo are expected 
to be nearly statistically isotropic, implying that the average ${\rm RM} = \langle \int\rho\BB\cdot d\bfl\rangle=0$. 
We therefore focus on the standard deviation of the RM, 
$\sigma_{\rm RM}$, particularly in the saturated state of the dynamo. 
It is also useful to normalize $\sigma_{\rm RM}$ by 
\EQA
\sigma_{\rm RM0} &=&  \frac{K}{\mu m_{p}} \frac{\meanrho B_{\rm rms}}{\sqrt{3}} L \sqrt{\frac{2\pi}{k_{\rm f}L}} 
\sim 444\,{\rm rad\,m^{-2}}\,\left(\frac{\overline{n}_{\rm e}}{1\, {\rm cm^{-3}}}\right) \nonumber \\
&&\times \left(\frac{B_{\rm rms}}{3 \mkG}\right)\,
\left(\frac{L}{1\,{\rm kpc}}\right)^{1/2}\,\left(\frac{l_{\rm f}}{100\,{\rm pc}}\right)^{1/2},
\label{sigmarm}
\ENA 
obtained from a simple model of random magnetic fields, where the fields are assumed to
be random with a correlation length $l_{\rm f} = 2\pi/k_{\rm f}$, in a box of length $L$ and uniform density
$\bar\rho$ (BS13). We then calculate the normalized standard deviation 
$\sigmarm=\sigma_{\rm RM}/\sigma_{\rm RM0}$ of the set ${\rm RM}(x_{i}, y_{i}, t)$. For LOS along 
$x$ and $y$, $B_{x}$ and $B_{y}$ values are to be used, respectively, in equation~(\ref{rmeq}). The final 
$\sigmarm$ is then taken to be the average of the $\sigmarm$ values obtained from the estimate of the 
standard deviations of ${\rm RM}$ along the three LOS. For a small-scale dynamo generated field, 
ordered on the forcing scale,we would expect $\sigmarm \sim 1$.
%%%%%%%%%%%%%%%%%%%%%%%%%%%%%%%%%%%%%%%%%%%%%%%%%%%%%%%%
\begin{figure}
\centering
\includegraphics[width=0.87\columnwidth]{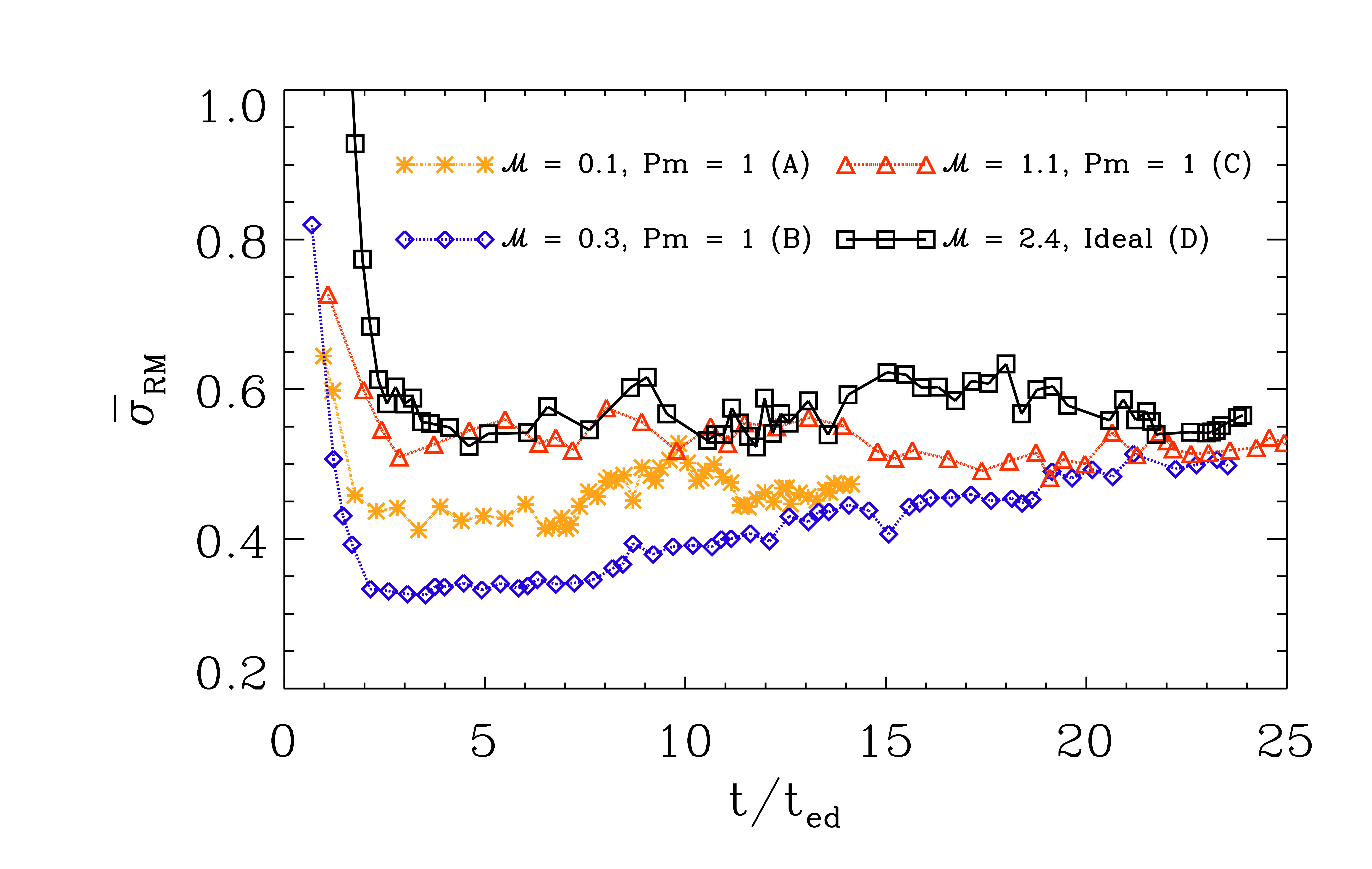}
\vspace{-1em}
\caption{Evolution of $\sigmarm$ with $t/t_{\rm ed}$ for all the runs listed in Table~\ref{sumsim}.
}
\label{sigrm_all}
\vspace{-1.0em}
\end{figure}
%%%%%%%%%%%%%%%%%%%%%%%%%%%%%%%%%%%%%%%%%%%%%%%%%%%%%%

\subsection{Evolution of $\sigmarm$, magnetic and velocity integral scales}

Recall that our seed magnetic field is characterized by a weak, large-scale vertical mean field. This 
implies that $\sigmarm$ would be initially dominated by the large-scale coherence of this vertical 
mean-field. Thereafter, due to turbulent driving the large-scale component would be randomly 
sheared leading to the emergence of random fields coherent on much smaller scales. Consistent 
with this argument, Fig.~\ref{sigrm_all} shows that there is initially a large $\sigmarm$ which 
subsequently decreases to a minimum. As the dynamo amplifies the fields, the field orders itself 
due to Lorentz forces. In most runs, we find that the steady-state $\sigmarm$ evolution curves are 
close to each other. Up to $\mathcal{M} = 2.4$, the last column in Table~2  shows that $\sigmarm$ in 
the saturated phases ranges between $0.46$ and  $0.55$. This indicates that the fields generated by the 
fluctuation dynamo in a variety of flows, have statistical properties that yield similar $\sigmarm$ in the 
saturated state. Remarkably, the value of $\sigmarm$ for $\mathcal{M} \approx 0.1$ 
obtained in run A is identical to that obtained from run F of BS13 who used a different code with 
delta-correlated turbulent driving and weak random initial seed fields.  
Thus the saturated state of the dynamo appears to be statistically independent
of the initial seed field configuration, for sufficiently weak seed fields.

To understand better the field coherence and its correlation to the $\sigmarm$, we directly compute 
the magnetic and the velocity integral scale from $M(k,t)$ and $K(k,t)$ as,
\EQ
l_{\rm int}^{M}(t) = \frac{\int ({2\pi/k})\,M(k,t)\,dk}{\int M(k,t)\,dk}, \,\,\,
l_{\rm int}^{V}(t) = \frac{\int ({2\pi/k})\,K(k,t)\,dk}{\int K(k,t)\,dk}.
\label{intscales}
\EN 
%%%%%%%%%%%%%%%%%%%%%%%%%%%%%%%%%%%%%%%%%%%%%%%%%%%%%
\begin{table}
\caption{Summary of the average values of : $l_{\rm int}^{V}$, $l_{\rm int}^{M}$, their ratio, and 
the $\sigmarm$ in the saturated state obtained in the different runs. *Run F corresponds to a 
$512^{3}$ non-ideal run from BS13 at $\mathcal{M} = 0.1$. 
}
\vspace{-0.5em}
\resizebox{0.5\textwidth}{!}{%
\begin{tabular}{|c|c|c|c|c|c|} \hline 
Run & $\mathcal{M}_{\rm rms}$ & Saturation & Saturation & Ratio of & Saturation \\
        & & $l_{\rm int}^{V}$ & $l_{\rm int}^{M}$ & $(l_{\rm int}^{V}/l_{\rm int}^{M})_{\rm sat}$ & 
        $\sigmarm$ \\ \hline 
A & $\approx 0.1$ & 4.4 & 1.6 & 2.75 & 0.46 \\ 
B & $\approx 0.3$ & 3.7 & 1.5 & 2.46 & 0.49\\ 
C & $\approx 1.1$ & 3.9 & 1.5 & 2.60 & 0.52\\ 
D & $\approx 2.4$ & 3.8 & 0.7 & 5.42 & 0.55 \\ 
F* & $\approx 0.1$ & 3.3 & 1.1 & 3.00 & 0.46 \\ \hline 
\end{tabular}
}
\label{table2}
\vspace{-1.2em}
\end{table}
%%%%%%%%%%%%%%%%%%%%%%%%%%%%%%%%%%%%%%%%%%%%%%%%%%%%%%%%%
Table~2 further shows that $l_{\rm int}^{M}$ values are similar in all the non-ideal runs,
with $l_{\rm int}^{V}/l_{\rm int}^{M} \sim 3$.
On the other hand, run D which has the highest $\sigmarm = 0.55$, has a smaller $l_{\rm int}^{M}=0.7$,
and larger $l_{\rm int}^{V}/l_{\rm int}^{M} \sim 5.4$.
This suggests that due to shock compression in supersonic turbulence, there must be significant 
contribution to $\sigmarm$ from high density regions with larger fields.
Altogether, except run D, the estimates of $l_{\rm int}^{V}/l_{\rm int}^{M}$ obtained in our 
non-ideal runs are similar to those obtained by BS13. 

\subsection{RM contributions from regions of different field strengths and densities}

The results from the previous subsection suggest that even in $\mathcal{M}\geq1$ cases, fluctuation dynamo 
fields are sufficiently coherent to obtain significant amount of RM. A key question is whether this coherence 
is arising from the strong fields or from the more volume filling less intense fields? 
Thus, we calculate the RM along each LOS for all the runs, by excluding regions where the field satisfies the 
constraint $B^{2} = (B_{x}^{2} + B_{y}^{2} + B_{z}^{2}) > (n\,B_{\rm rms})^{2}$, with $n=1$ and $2$. In subsonic 
flows, we find that the strong field regions (i.e., with $n = 2$) contribute only about 20 per cent to the RM 
in agreement with BS13. We extend the analysis to magnetic fields in transonic and supersonic flows.
%%%%%%%%%%%%%%%%%%%%%%%%%%%%%%%%%%%%%%%%%%%%%%%%%%%%%%%%
\begin{figure}
\centering
\includegraphics[width=0.41\textwidth]{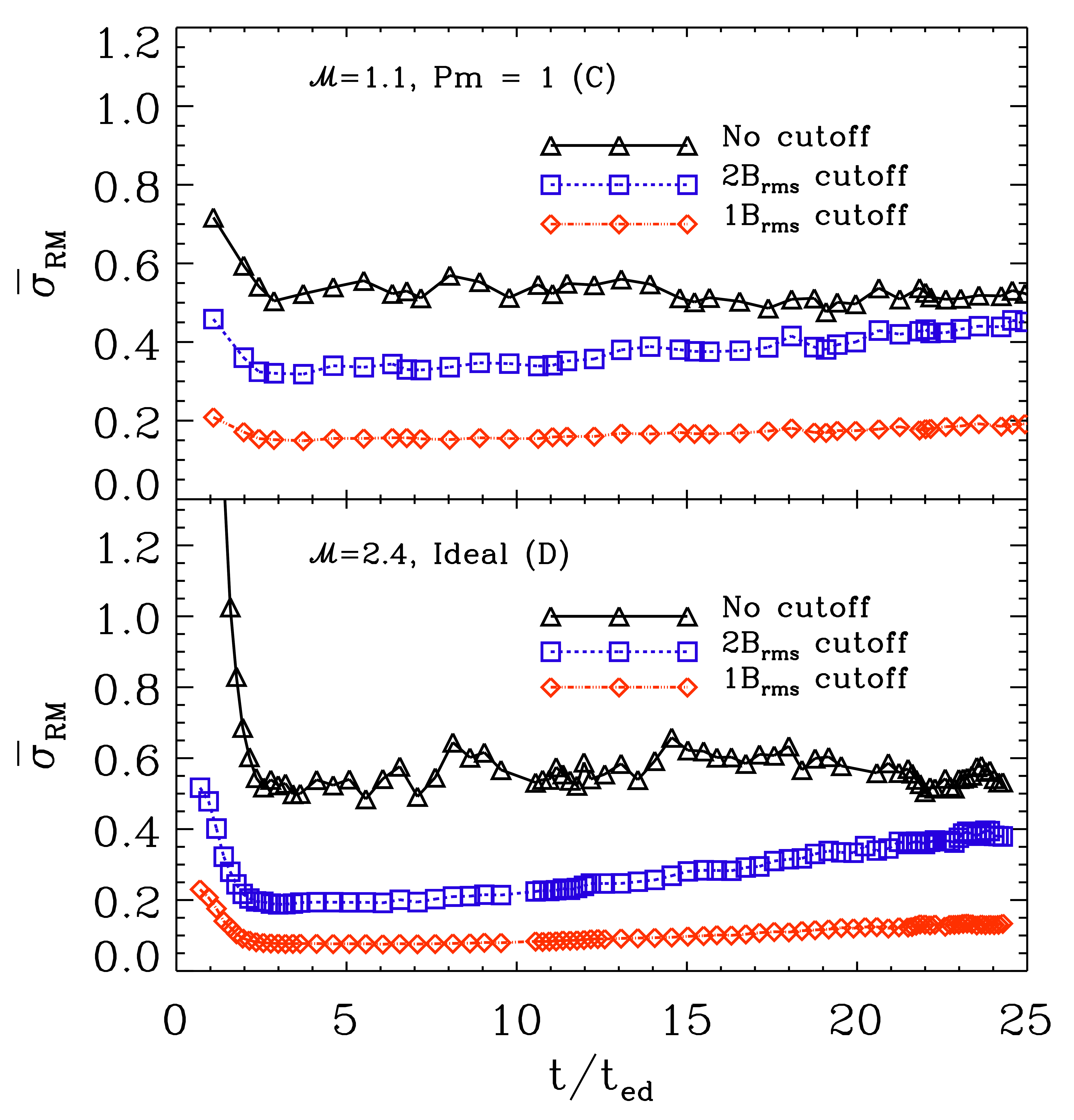}
\vspace{-1em}
\caption{Evolution of the normalized RM $(\sigmarm)$ for runs C (top panel) and D (bottom panel), 
determined excluding the regions with $|\BB| > n\,B_{\rm rms}$. The squares show the result of 
excluding $|\BB| > 2\,B_{\rm rms}$, while the diamonds show the result of excluding regions with 
$|\BB| > B_{\rm rms}$. 
}
\label{bcuts}
\vspace{-1.5em}
\end{figure}
%%%%%%%%%%%%%%%%%%%%%%%%%%%%%%%%%%%%%%%%%%%%%%%%%%%%%%

\begin{figure}
\centering
\includegraphics[width=0.41\textwidth]{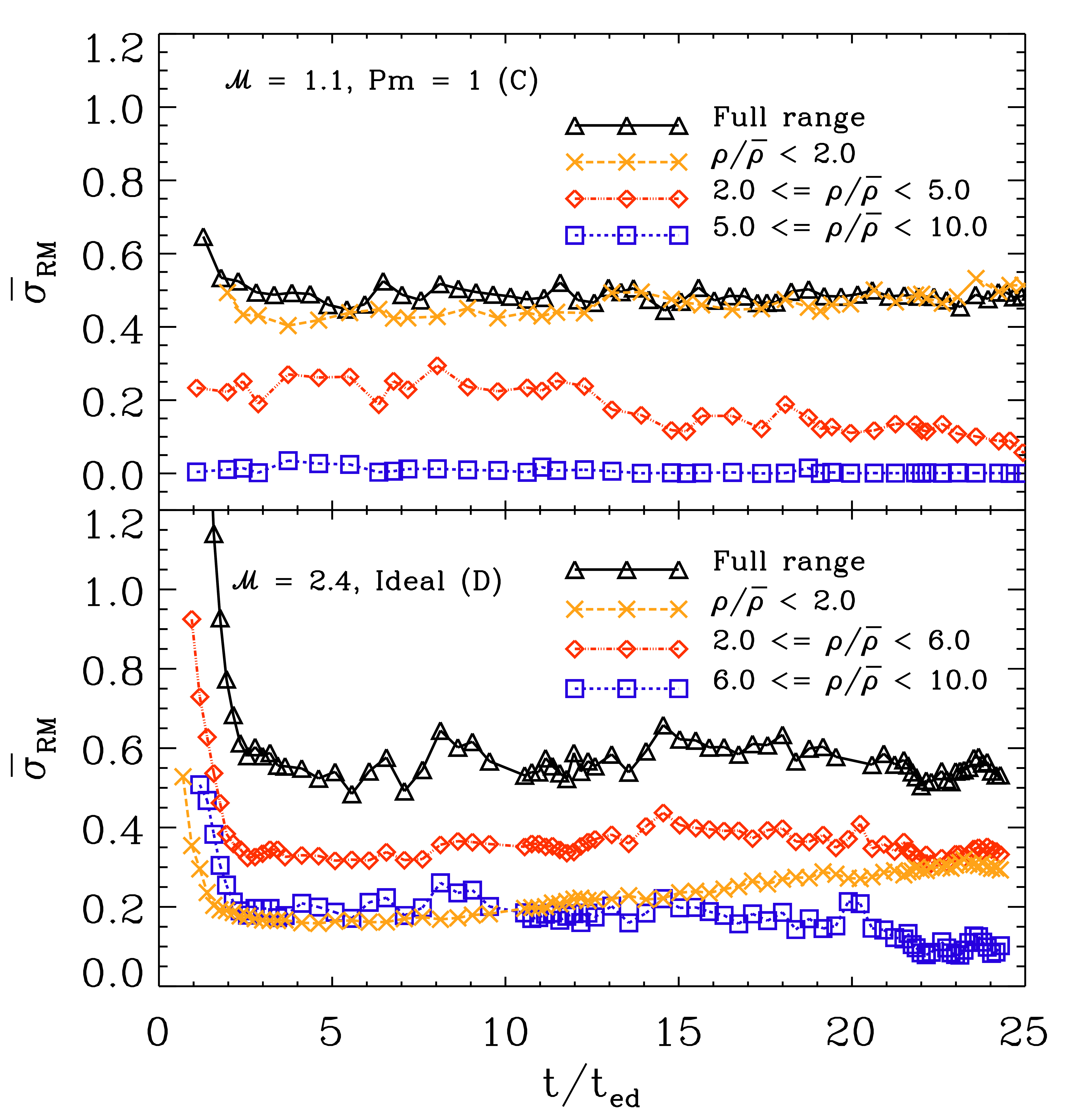}
\vspace{-1em}
\caption{Evolution of the normalized RM $(\sigmarm)$ for runs C (top panel) and D (bottom panel), 
determined from regions of different density ranges.}
\label{denscuts}
\vspace{-1.5em}
\end{figure}

Results for run C (top panel) and run D (bottom panel) are shown in Fig.~\ref{bcuts}. Black solid lines with 
triangles in both panels show the $\sigmarm$ obtained without introducing any cutoffs. 
For both runs, $\sigmarm$ drastically reduces if one excludes regions with field strengths $B > B_{\rm rms}$ 
(red dash-dotted lines with diamonds). 
The reduction in $\sigmarm$ when one removes regions with field strength $B > 2\,B_{\rm rms}$
is still significant for both runs in the growth phase. However, at saturation the reduction is $\leq 10$ per cent 
for the transonic case and $\sim 30$ per cent in the supersonic case. 
Thus for subsonic and transonic driving most of the RM contribution at saturation is dominated by the 
general sea of volume filling fields similar to the earlier results of BS13. 
However in Run D (corresponding to supersonic case), there is still a significant contribution to RM from 
high field regions initially, which then decreases on saturation. 
The density enhancement in such supersonic turbulence could then be playing a role. This can in fact be 
seen in Fig.~\ref{denscuts} where we show the evolution of $\sigmarm$ contributed by regions of different 
over density ranges, instead of cutoffs. We see that while in the transonic case high density regions do not 
contribute significantly to $\sigmarm$, in the supersonic case (Run D) moderate overdense regions
with $2 \leq \rho/\bar\rho < 6.0$, around $\sim \mathcal{M}^2$ do significantly contribute at all times. 
This includes the earlier times, when the contribution to $\sigmarm$ is dominated by $B > 2\,B_{\rm rms}$
regions as in Fig.~\ref{bcuts}.  But later on, the contribution from $\rho/\bar\rho < 2.0$ 
also increases as the field in Run D saturates, simultaneously as contribution from $B > 2\,B_{\rm rms}$ 
in Fig.~\ref{bcuts} decreases. Thus such lower overdensity range perhaps coincides with the lower 
field strength regions. 

%%%%%%%%%%%%%%%%%%%%%%%%%%%%%%%%%%%%%%%%%%%%%%%%%%%%%%%%
%%%%%%%%%%%%%%%%%%%%%%%%%%%%%%%%%%%%%%%%%%%%%%%%%%%%%%

\section{Conclusions}
\label{conc}

The prime objective of this work was to explore Faraday RM and the degree of coherence 
arising from magnetic fields generated by fluctuation dynamos in young galaxies at high 
redshifts. Since the ISM turbulence of young galaxies is expected to be supersonic, addressing the 
stated goals necessitated the need for simulating fluctuation dynamos in such turbulent flows. 
Equally, it is also imperative to compare our findings with previously known results from 
fluctuation dynamo simulations in subsonic flows. To this effect, we focused on fluctuation dynamo 
action in turbulent flows consisting of both ideal and non-ideal runs (at $\Pm=1$) for a range of 
Mach numbers $\mathcal{M} = 0.1 - 2.4$. 

By shooting $3N^{2}$ LOS through the simulation box we directly computed the standard deviation of the 
measured RMs, $\sigmarm$. Our results show (see Fig.~\ref{sigrm_all}) that up to $\mathcal{M} = 2.4$ 
probed here $\sigmarm = 0.46 - 0.55$ at saturation, independent of the Mach number of the flow. 
Remarkably, the estimates of $\sigmarm$ for transonic and supersonic runs are similar to the values 
obtained in a previous work of BS13 for subsonic turbulence. From equation (\ref{sigmarm}) such values 
of $\sigmarm$ lead to a random ${\rm RM} \sim 16 - 48$ rad m$^{-2}$ in the galactic context 
(for $\overline{n}_{\rm e} = 0.1 - 0.3\,{\rm cm^{-3}}, B_{\rm rms} = 3.0 \mkG, L = 500\,{\rm pc}, 
l_{\rm f} = 100\,{\rm pc}$ and $\sigmarm = 0.5$) consistent with the observations of \citet{Farnes+14}. 
We also computed the evolution of the integral scale of the field and flow. By comparing 
the integral scale of the field, $l_{\rm int}^{M}$ with $\sigmarm$, we find that there is a reasonable match 
with the expected $\sigmarm \propto \sqrt{l_{\rm int}^{M}}$ scaling in the subsonic cases. 
However the supersonic case, which has a higher $\sigmarm$, showed lower field coherence scale 
$l_{\rm int}^{M}$ compared to the subsonic cases, indicating a possibly significant contribution to 
$\sigmarm$ from overdense, large field but small scale regions. 
We also find that RM does not decrease substantially if one removes regions with field strengths larger 
than $2\,B_{\rm rms}$ in the subsonic cases implying that it is the general sea of 'volume filling' fields 
that crucially determine the RM, rather than the strong field regions. 
However, for dynamo generated field in supersonic flows, strong field regions and those with 
overdensities up to $\rho/\bar\rho \sim \mathcal{M}^2$ also contribute significantly to RM.

In the near future, we plan to extend our analysis to high $\Pm$ systems with resolutions sufficient for 
resolving large $\Rey, \Rm$ values. It will also be worthwhile to explore other observables such as the 
synchrotron emissivity and polarization, indispensable for weaving a coherent picture of magnetic fields 
in young galaxies.  

\section*{Acknowledgements}
Sharanya Sur acknowledges computing time at CDAC National Param Supercomputing Facility Pune, 
Texas Advanced Computing center and the use of HPC facilities at IIA. Pallavi Bhat is supported 
by NSF-DOE Award No. DE-SC0016215. The software used in this work was in part developed by 
the DOE NNSA-ASC OASCR Flash Center at the University of Chicago.

\bibliographystyle{mnras}
\bibliography{ssdyn_ssurbib}

\label{lastpage}

\end{document}